\documentclass[letterpaper,12pt]{article} 

\usepackage{temp}

\usepackage{amsmath,amssymb}

\usepackage{cite}
\usepackage{jabbrv}
\RequirePackage{geometry}
\usepackage{color}
\usepackage{graphicx}
\usepackage{mathptmx,courier,textcomp}
\usepackage{helvet}

\begin{document}

\title{Simultaneous Existence of Ultra-high and Ultra-low Spectral Sensitivity for Directional Couplers}

\author{Garima Bawa$^1$, Indrajeet Kumar$^2$, Saurabh Mani Tripathi$^{1,2}$}

\address{$^1$Department of Physics, Indian Institute of Technology Kanpur, Kanpur 208016\\
$^2$Centre for Lasers and Photonics, Indian Institute of Technology Kanpur, Kanpur 208016}
\email{garimab@iitk.ac.in, ikumar@iitk.ac.in, smt@iitk.ac.in}

\vskip 1cm
\begin{center}
	\textbf{Abstract}\\
\end{center}
We present the experimental evidence for the existence of two exciting wavelengths, termed $critical$ and $cross-over$ wavelength in the transmission spectrum of fiber-optic directional coupler, whose properties are conjugate to each other. The spectral shift associated with the transmission maxima/minima suddenly flip around these wavelengths and the spectrum shows largest (nil) spectral shift for the transmission maxima/minima closest to the $critical$ ( at the $cross-over$) wavelength, corresponding to the same perturbation parameter. A theoretical explanation of the observed experimental behavior has also been presented, highlighting that the underlying mechanisms for the existence of these wavelengths are entirely different. The knowledge of the precise spectral location of these wavelengths is necessary to avoid false alarms.

\section{Introduction}
Spectral and power interrogation techniques are among the most widely used interrogation techniques employed in photonic systems using modal coupling in optical waveguides. In the spectral interrogation technique, the wavelength shift associated with the transmission (reflection) reference maxima/minima are measured with respect to the changes in the external perturbation parameters. In contrast, in the power interrogation technique, the changes in the external perturbation parameters are quantified in terms of the changes in the power level recorded at the fixed reference maxima/minima. Both of these techniques have their inherent advantages and drawbacks. Spectral interrogation technique, for example, is immensely accurate and is immune to input power fluctuations and/or connector losses to which the power interrogation technique suffers badly. The power interrogation technique, on the other hand, is a much cheaper alternative and does not require expensive optical spectrum analyzer and broadband sources. It is the accuracy of the detection system, and absence of false alarms, which dictates the choice of the detection technique,
and therefore, spectral interrogation technique is almost always preferred over the power interrogation technique. A number of highly sensitive sensors for detection of a variety of bio \cite{Dandapat}, physical \cite{Huang} and chemical \cite{Aray} parameters, based on spectral interrogation technique have been reported. Although offering extremely high sensitivity, these sensors are also prone to false alarms arising due to erroneous association of a particular type of wavelength shift (blue or red) with increase/decrease in measurand quantity whereas the sensor records it conversely. For example, the experimental findings reported by different groups show contrasting results (entirely opposite nature of wavelength shift) with respect to the same perturbation applied to the sensor. 
Employing directional couplers (DCs) fabricated using SMF-28 (Corning, New York, 14831, USA), Shanshan et al. (- 1.13 nm/$^\circ C$ over 17$ ^\circ C$ - 31.6$ ^\circ C$) \cite{Wang} and Yuxuan et al. (- 5.3 nm/$^\circ C$ over 35$ ^\circ C$ - 45$ ^\circ C$) \cite{Jiang} have reported a negative spectral shift  with increase in ambient temperature while Ming et al. (+ 11.96 pm/$^\circ C$ over 247$ ^\circ C$ - 1283$ ^\circ C$ ) \cite{Ding} and Pengfei et al. (+ 36.59 pm/$^\circ C$ over 700$ ^\circ C$ - 1000$ ^\circ C$) \cite{Wang2} have reported a positive spectral shift. However, no explanation of the opposite spectral shifts is available.
Similar contrasting results have also been reported for multimodal-interference-effect based sensors \cite{SMT1,Garima1,Garima2,Talata,Li}.

In this letter, we explain this ambiguity by demonstrating the existence of two unique wavelengths in the transmission spectrum of a DC, whose properties are conjugate to each other. These wavelengths are termed as $critical$ and $cross-over$ wavelengths. The spectral shift in the transmission spectrum suddenly flip around these wavelengths and the spectrum shows largest (nil) spectral shift for the transmission maxima/minima closest to the $critical$ (at the $cross-over$) wavelength, corresponding to the same perturbation parameter. A theoretical explanation of the observed experimental behavior has also been presented, along with an explanation of the origin of these unique wavelengths.

\section{Experimental Details}
\begin{figure}[t!]
	\begin{center}
		\includegraphics [width=10cm]{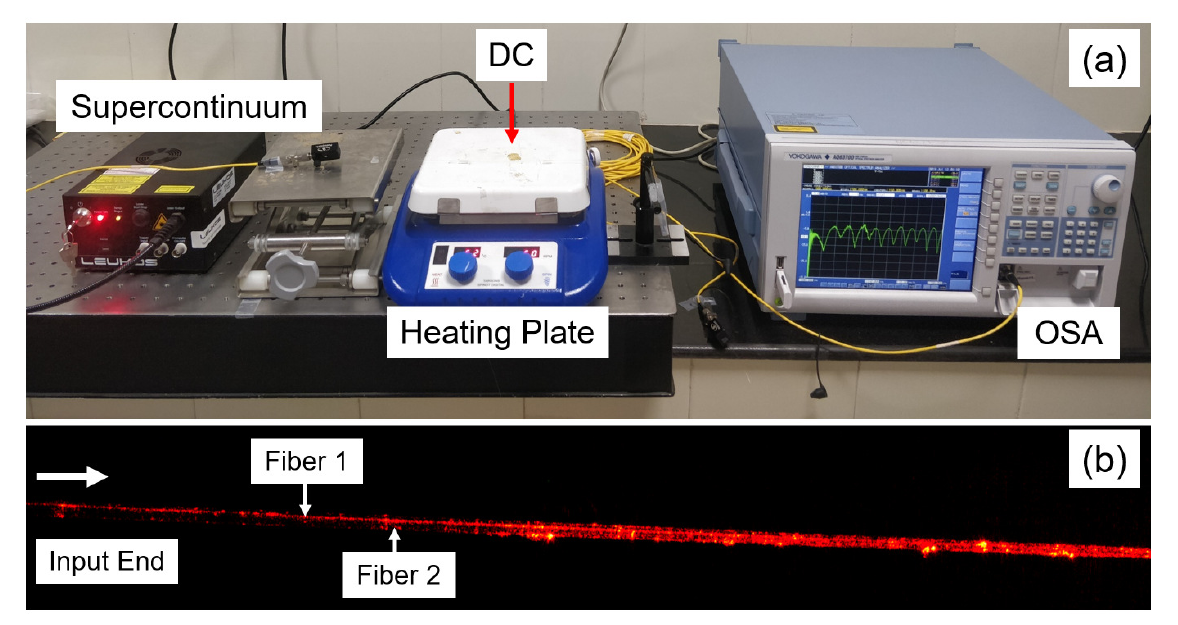}
	\end{center}
	\caption {(a) Photograph of the experimental setup, and (b) optical microscopic image of the fabricated DC.}
	\label{setup}
\end{figure}

In our experiments, we fabricated several DCs using the heat and fuse method \cite{Ghatak1}. Small sections ($\sim$7 cm) of single-mode optical fibers were unjacketed and properly cleansed in ethanol ($\geqslant$99.8$\%$ (GC), Merck Millipore); after that, the cleansed portions were twisted together. To ensure a uniform elongation, a pre-calibrated dynamic tension (7 - 22mN) was applied to both the fibers; this also prevents any kink in the twisted region while fabricating the DC. The twisted portion was then heated using a butane burner set up for 8-10 minutes with optimized fiber to flame separation, blow rate, and angle of gas flow to get a continuous softening of the fiber glass without melting the fibers.  
The light was launched into one of the fiber using a supercontinuum source (LEUKOS SM-30-450) and the transmission spectra through the cross port were continuously monitored using an optical spectrum analyzer (YOKOGAWA AQ6370D). Fiber fusion was stopped once a power loss of $\sim$15 dB was achieved at various minima across the transmission spectrum. The experimental setup and optical microscopic image of the fabricated DC are shown in Figs. \ref{setup}($a$), and \ref{setup}($b$), respectively. 

\section{Results and Discussion}
The temperature response of the fabricated DCs was studied by heating the DC through a heating plate. To maintain a uniform temperature throughout the coupling region and stabilize its transmission response, the couplers were kept at fixed temperatures for $\sim$20 min before making observations. Similar to the fabrication process, bends across the coupling region were avoided by maintaining constant tension throughout the experiments by fixing the fibers to a stationary stage near launching end and applying a fixed force near the other end.

\begin{figure}[b!]
	\begin{center}
		\includegraphics [width=10cm]{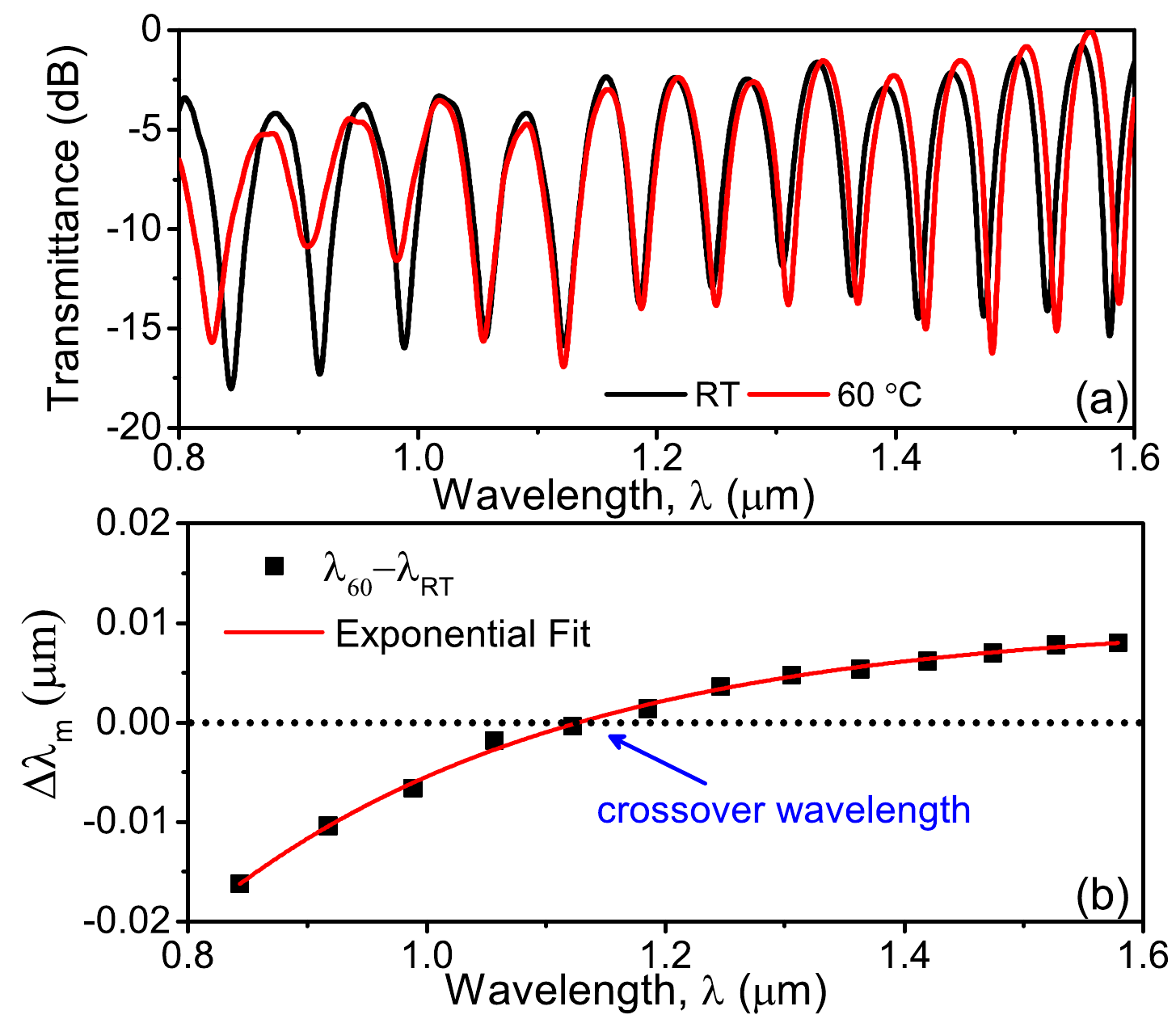}
	\end{center}
	\caption {($a$) Experimental transmission spectra and ($b$) Spectral shifts corresponding to various transmission minima of the DC employing two identical fibers (Corning SMF-28) at T = $25\ ^\circ C$ (RT), and $60\ ^\circ C$.  
		Cross-over wavelength is found to exist at 1.167 $\mu$m.}
	\label{cross-over}
\end{figure}

The transmission spectra were recorded at various temperatures starting from room temperature (RT) i.e., 25 $^\circ C$ till 60$^\circ C$ with an interval of 5 $^\circ C$, for a DC fabricated using two identical single-mode optical fibers (Corning SMF-28, New York, 14831, USA). To clearly depict the spectral shifts in Fig. \ref{cross-over}($a$) we have plotted the transmission spectra recorded at two extreme temperatures T = 25 $^\circ C$ (RT) (black curve) and T = 60 $^\circ C$ (red curve). The interaction length of the coupler was $\sim$8 mm and coupler width was 9.8$\pm$2 $\mu m$. We observe the existence of a point of inflection, which we term as $cross-over$ wavelength, at 1.167 $\mu$m: maxima/minima lower to this wavelength show a blue spectral shift with increasing temperature whereas maxima/minima on the higher side of $cross-over$ wavelength show a red spectral shift with increasing temperature. This explains why experimental results reported by different groups showed contrasting results (entirely opposite nature of wavelength shift) with respect to the same perturbation applied to the sensor \cite{Wang, Jiang, Wang2, Ding}. In all probability the results reported by different groups were carried out with their reference wavelengths lying on different sides of the cross-over wavelength, e.g., in \cite{Wang, Jiang} the experiments were carried out near 1470 nm whereas in \cite{Ding} the experiments were carried out around a reference wavelength of 1220 nm.
More interesting observations are made once we plot the spectral shifts corresponding to various reference maxima/minima ($\Delta\lambda_m$) in Fig. \ref{cross-over}($b$). It shows ($i$) an exponential dependence of $\Delta\lambda_m$ on the location of the reference maxima/minima, and ($ii$) spectral shifts are of opposite nature around $cross-over$ wavelength and increase further as we move away from the $cross-over$ wavelength with $\Delta\lambda_m$ = 0 at 1.167 $\mu$m ($cross-over$ wavelength). The exponential dependence of the spectral shift, and hence the sensitivity (= \begin{math} \frac{d\lambda_m}{d\chi} \end{math}, $\chi$ is the perturbation parameter), on the location of maxima/minima suggests the existence of a $critical$ wavelength on the lower wavelength side similar to the one observed in the transmission spectrum of single-multi-single mode structures \cite{SMT1}. The $critical$ wavelength for this structure, however, does not fall in the observable wavelength range (600 nm - 1700 nm) of the optical spectrum analyzers used in our experiments. Therefore, we could not record direct evidence of $critical$ wavelength. However, the trails of it were quite nicely captured by our experiment in terms of the exponentially increasing spectral shifts (0.45 $ nm / ^{\circ} C$ at the lowest minima). 

\begin{figure*}
	\centering
	\includegraphics [width=15cm]{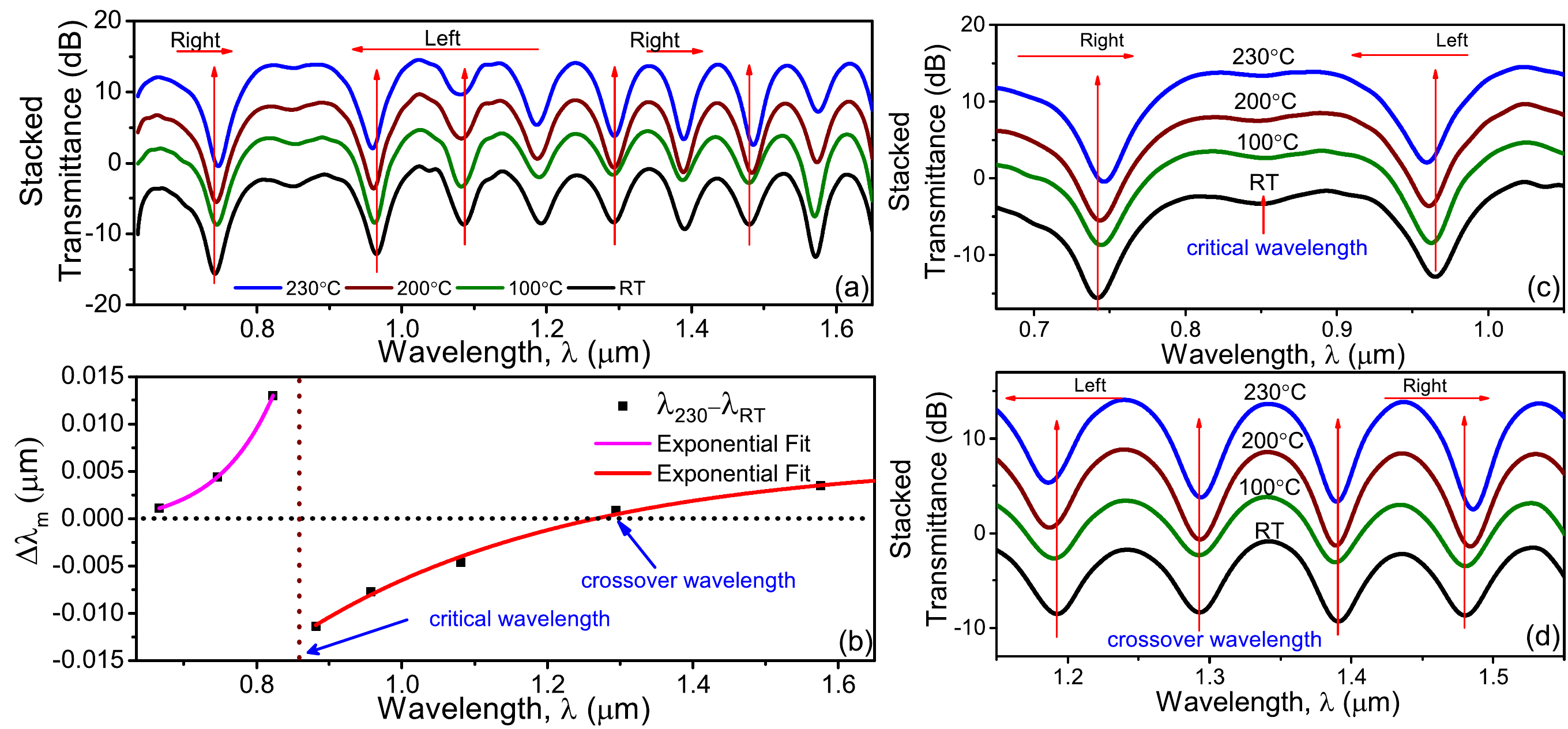}
	\caption{ ($a$) Experimental transmission spectra and ($b$) Spectral shifts corresponding to various transmission minima of the DC employing two non-identical fibers (Corning SMF-28 and Fibercore$^{TM}$ PS1250/1500 photosensitive fiber). Zoomed transmission spectra depicting ($c$) $critical$ wavelength at 0.850 $\mu$m, and ($d$) $cross-over$ wavelength at 1.266 $\mu$m at T = $25\ ^\circ C$ (RT), $100\ ^\circ C$, $200\ ^\circ C$ and  $230\ ^\circ C$ .  \textbf{(The transmission spectra have been stacked with +5 dB offset.)}}
	\label{crit-cross}
\end{figure*}

In order to get direct evidence of the $critical$ wavelength, which corresponds to maximum sensitivity, we fabricated several DCs with different interaction lengths using identical fibers: Corning SMF-28 or Fibercore$^{TM}$ PS1250/1500 photosensitive fiber, and a combination of these. Using identical fibers we observed a similar behavior of the exponential dependence of spectral shift on the spectral location of various maxima/minima in the transmission spectrum. Interestingly, however, using DCs fabricated with non-identical fibers (Fig. \ref{crit-cross}(a)) we observed the existence of $critical$ wavelength at 0.850 $\mu$m whose properties are conjugate to $cross-over$ wavelength observed at 1.266 $\mu$m for the same DC. 
To illustrate this further, we have plotted the spectral shift versus the position of the reference maxima/minima in Fig. \ref{crit-cross}(b). This figure clearly shows that while $\Delta\lambda_m$ is largest for the maxima/minima around the $critical$ wavelength, it is zero for maxima/minima at the $cross-over$ wavelength. Also, nature of the flip of spectral shifts around the $critical$ and $cross-over$ wavelengths are opposite to each other. We would like to mention here that the spectral shifts (54.56 pm/$^{\circ}C$ considering the first minima on both sides together) observed for this structure are not large even for sufficiently large temperature variation ($\sim$ 230 $^\circ C$), which is mainly due to the use of non-identical fibers with dopants (GeO$_2$ and B$_2$O$_3$) of opposite thermo-optic coefficients ($dn/dt$ $>$ 0 for GeO$_2$ and $dn/dt$ $<$ 0 for B$_2$O$_3$) constituting the DC.

\begin{figure}[b!]
	\begin{center}
		\includegraphics [width=10 cm]{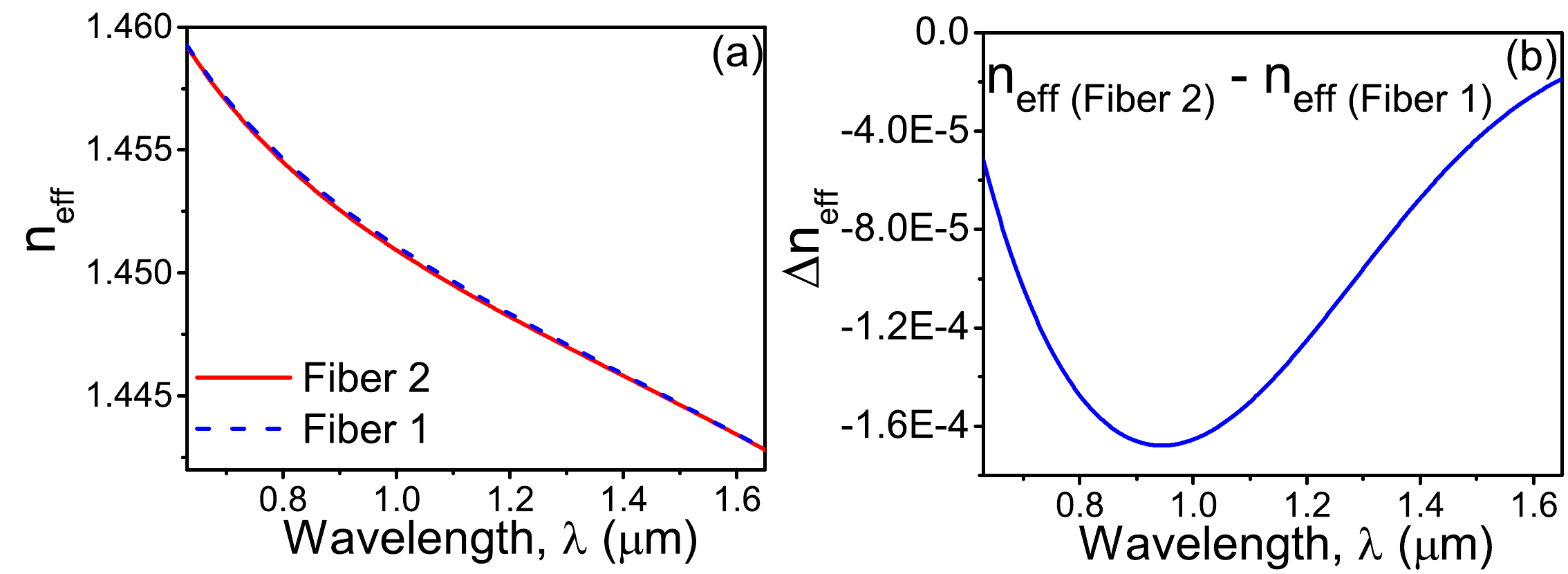}
	\end{center}
	\caption{Spectral Variation of  ($ a $) effective indices ($n_{eff}$) of fundamental mode of Fiber 1 and Fiber 2 and ($ b $) difference in effective indices of the fundamental modes of the two fibers. }
	\label{neff}
\end{figure}

To understand the origin of $cross-over$ $\&$ $critical$ wavelengths and their conjugate properties theoretically, we consider a DC consisting of two optical fibers (core radii being 1.75 $\mu$m over the coupler region) kept very close to each other (center to center separation of 4.2 $\mu$m), such that the evanescent fields of their propagating modes overlap, causing a periodic energy exchange between the fibers. Since the exact dopant and their concentrations in the core region of the optical fibers used in our experiments were not disclosed by the manufacturers, in our simulations we consider the cladding region of fibers to be made of fused SiO$_2$, and the core region of Corning SMF-28 made of 3.1 mole$\%$ GeO$_2$ in SiO$_2$ (henceforth Fiber-1) and that of photosensitive Fibercore$^{TM}$ 1250/1500 made of 4.03 mole$\%$ GeO$_2$ and 9.7 mole$\%$ B$_2$O$_3$ in SiO$_2$ host (henceforth Fiber-2). To check the proximity of the opto-geometric parameters used in our simulations with the experimental fibers we simulated the mode field diameter (MFD) at 1550 nm using the aforementioned doping concentrations and found an excellent agreement with the MFD of experimental fibers. We further, found that the effective indices of the fundamental mode (HE$_{11}$) for both the fibers are very close to each other, matching till the fourth decimal place, as has been shown in Fig. \ref{neff}(a) and \ref{neff}(b). The modal field distribution of the fundamental mode supported by these fibers are therefore nearly identical to each other. For ease of calculation of the coupling coefficient ($\kappa$), we have used an analytical expression employing identical fibers given below \cite{Ghatak1} \cite{Snyder}
\begin{equation}
\kappa(d)=\dfrac{\lambda_{0}}{2\pi n_{1}}\dfrac{U^2}{a^2 V^2} \dfrac{K_{0}(Wd/a)}{K^{2}_{1}(W)}.
\label{6}
\end{equation}
Here $\lambda_{0}$ is the free space wavelength, $n_1$ is the core refractive index, $a$ is the core radius, $d$ is the separation between fiber axes, and $K_\nu (x)$ represents the modified Bessel function of order $\nu$. $V$, $U$, and $W$ have their usual meaning.
The periodic powers carried by the first ($ \mid a(L) \mid^2$) and the second ($ \mid b(L) \mid^2$) SMFs of the DC fabricated using identical fibers are given by \cite{Ghatak2}
\begin{equation}
\mid a(L)\mid^2=1-sin^2(\kappa L)\ \ \ ;\ \ \ \mid b(L)\mid^2=sin^2( \kappa L)
\end{equation}    

In our study, Sellmeier relation \cite{Adams} has been used to incorporate the wavelength dependence of the refractive indices of the fibers. Temperature-dependent refractive indices of the core, as well as cladding regions, have been obtained using the relation $n=n_{0}+(dn/dT)(T-T_0)$, where $n_{0}$ is the refractive index at room temperature $T_0$. The thermo-optic coefficient $(dn/dT)$ for fused silica, 100 $\%$ GeO$_2$ and 100 $\%$ B$_2$O$_3$ are known to be 1.06 $ \times $ 10$^{-5}/^{\circ}C$  \cite{SMT2}, 1.94 $ \times $ 10$^{-5}/^{\circ}C$, and -3.5  $ \times $ 10$^{-5}/^{\circ}C$, respectively. Since the core regions of the SMFs are not heavily doped, considering a linear dependence of $dn/dT$ on the respective concentrations \cite{thermo}, the value of $dn/dT$ for 3.1 mole$\%$ GeO$_2$ in SiO$_2$ was calculated to be 1.0972 $ \times $ 10$^{-5}/^{\circ}C$. The change in the core radius, a of the SMF and the interaction length, L of the DC with temperature were obtained using $\Delta a = \alpha a \Delta T$ and  $\Delta L = \alpha L \Delta T$ \cite{SMT2}. 	

\begin{figure*}
	\begin{center}
		\includegraphics [width=16 cm]{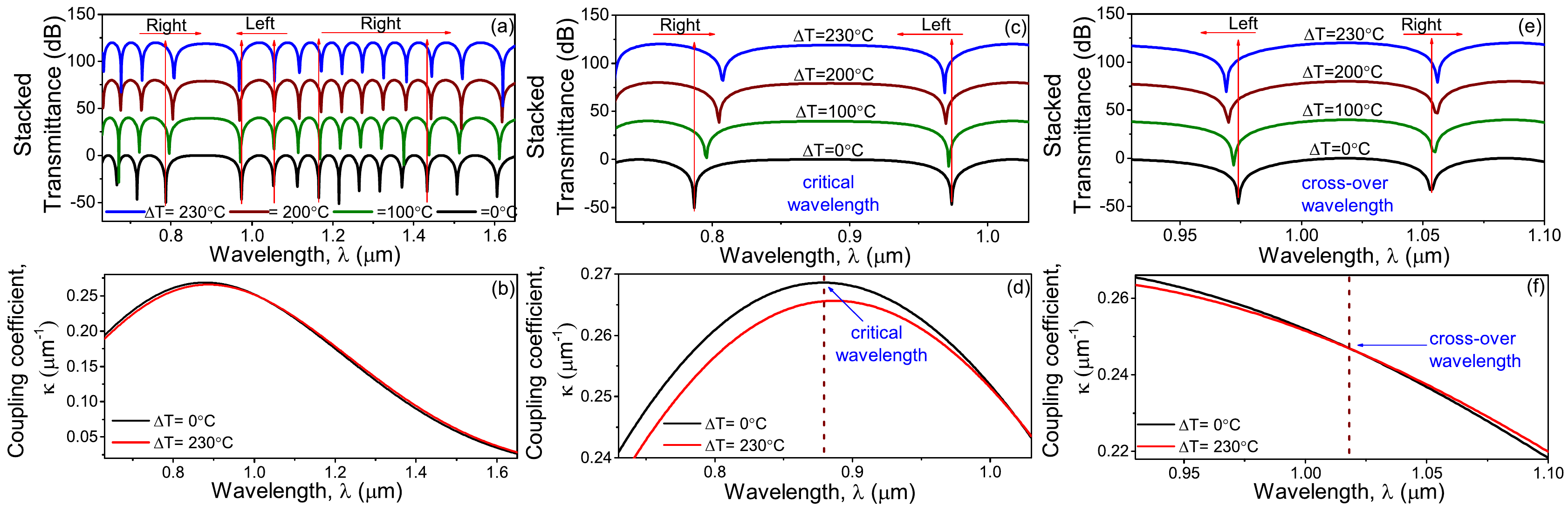}
	\end{center}
	\caption {($ a $) Theoretical transmission spectra for $\Delta T=0^\circ C$, $100\ ^\circ C$, $200\ ^\circ C$ and $230\ ^\circ C$ and ($ b $) Spectral variation of $\kappa$ for $\Delta T=0\ ^\circ C$, and $230\ ^\circ C$ over a wavelength range 0.63 - 1.65 $\mu m$ ; Zoomed ($ c $) Transmission spectra and ($ d $) Spectral variation of $\kappa$ over a wavelength range 0.73 - 1.05 $\mu m$ depicting the $ critical $ wavelength at 0.879 $\mu m$ and ; ($ e $) Transmission spectra and ($ f $) Spectral variation of $\kappa$ over a wavelength range 0.93 - 1.1 $\mu m$ depicting the $ cross-over $ wavelength at 1.034 $\mu m$. \textbf{(The transmission spectra have been stacked with +40 dB offset.)}}
	\label{simulated transmission}
\end{figure*}

In our simulations, we have fixed the interaction length equal to the tenth coupling length i.e., 140 $\mu m$ at $\lambda =$ 1.55 $\mu m$; and studied the transmission spectra at various temperature differences. The transmission spectra at four different temperature variations i.e. $\Delta T=0\ ^\circ C$, $100\ ^\circ C$, $200\ ^\circ C$ and $230\ ^\circ C$ are shown in Fig. \ref{simulated transmission}($ a $). From Fig. \ref{simulated transmission}($ a $), we observe that around the wavelength 0.879 $\mu m$ the transmission maxima/minima show ($i$) opposite spectral shifts with an increase in temperature, and ($ii$) the spectral shift of the maximas/minimas closest to it are largest. Both of these properties correspond to the $critical$ wavelength observed in our experiments. From Fig. \ref{simulated transmission}($a$) we also observe that around 1.034 $\mu$m, the spectrum once again flips its nature of spectral shift similar to our observations of $cross-over$ wavelength.

The overall transmission of the sensor depends upon the phase, $\phi=\kappa L$ supported by the coupler. To explain these peculiar behaviors, we first note that the variation in the phase, $\phi$ obtained through the Taylor series expansion, which, retaining the first-order terms only, gives
\begin{equation}
\Delta \phi = \dfrac{\partial \phi}{\partial \lambda} \Delta \lambda + \dfrac{\partial\phi}{\partial T} \Delta T .
\label{8}
\end{equation}

Unlike $L$, $\kappa$ is a function of both the operating wavelength ($\lambda$) as well as the perturbation parameter (here temperature $T$). For constant phase points ( i.e. $\Delta \phi =0$) we get
\begin{equation}
\frac{\Delta  \lambda}{\Delta T} = -\dfrac{\partial \phi}{\partial T}\dfrac{1}{L} \Big(\dfrac{\partial\kappa}{\partial \lambda}\Big)^{-1} \\
=-\Big[\dfrac{\partial \kappa}{\partial T} + \kappa \alpha \Big] \Big(\dfrac{\partial\kappa}{\partial \lambda}\Big)^{-1}
\label{11}
\end{equation}

Since the thermal expansion coefficient $(\alpha)$ and the coupling coefficient $(\kappa)$ are essentially positive quantities, the nature of the spectral shift of the transmission spectrum will be governed by $\dfrac{\partial \kappa}{\partial T}$ and $\dfrac{\partial \kappa}{\partial \lambda}$. In Fig. \ref{simulated transmission}($b$), we have plotted the spectral variation of the coupling coefficient, $\kappa$, over a wavelength range of 0.63 - 1.65 $\mu m$ at two different temperatures. The zoomed transmission spectra and the corresponding spectral variation of the coupling coefficient, $\kappa$, over a wavelength range of 0.75 - 1.05 $\mu m$ have been plotted in Fig. \ref{simulated transmission}($c$) and Fig. \ref{simulated transmission}($d$), respectively, clearly showing the existence of $critical$ wavelength at 0.879 $\mu m$. Fig. \ref{simulated transmission}($d$) shows that while $\dfrac{\partial \kappa}{\partial T}$ is monotonically negative throughout this wavelength range, $\dfrac{\partial \kappa}{\partial \lambda}$ is positive for $\lambda < 0.879$ $\mu m$ whereas it is negative for $\lambda > 0.879$ $\mu m$. Therefore, as per equation (\ref{11}) the transmission spectrum should show a red spectral shift for $\lambda < 0.879$ $\mu m$ and a blue spectral shift for $\lambda > 0.879 $ $\mu m$. This agrees excellently with nature of the spectral shift shown in Fig. \ref{simulated transmission}($c$) and the experimental behavior observed in our experiment (Fig. \ref{crit-cross} (c)).
Similarly, the spectral variation of $\kappa$ over a shorter wavelength range of 0.9-1.1 $\mu m$ is plotted in Fig. \ref{simulated transmission}($f$). We see that in contrast to Fig. \ref{simulated transmission}($d$), the slope $\dfrac{\partial \kappa}{\partial \lambda}$ is negative throughout this wavelength range whereas $\dfrac{\partial \kappa}{\partial T}$ is negative for $\lambda < 1.034$ $\mu m$ and positive for $\lambda > 1.034$ $\mu m$. Thus, as per equation (\ref{11}), we should get opposite spectral shifts around this wavelength, with lower wavelength side showing a blue spectral shift with increasing temperature. 
This again agrees excellently with the spectral shifts of the transmission spectrum plotted in Fig. \ref{simulated transmission}($e$) and the experimental behavior observed in our experiment (Fig. \ref{crit-cross} (d)).

We would like to mention here that while the flip of the spectral shift is common for both; the $critical$ as well as $cross-over$ wavelength, the physical mechanism responsible for the flip is entirely different for both the cases. It is also interesting to note that one wavelength corresponds to ultra-high sensitivity ($critical$ wavelength) whereas the other with ultra-low sensitivity ($cross-over$ wavelength) towards temperature.	

\section{Conclusion}
In conclusion, we have demonstrated and explained the existence of two unique wavelengths -- the $critical$ and the $cross-over$ wavelengths in the transmission spectrum of a DC, whose properties are conjugate to each other. The $critical$ wavelength has largest and opposite spectral shifts for the transmission maxima/minima closest to and on either side of it, whereas the $cross-over$ wavelength corresponds to zero spectral shift. We have also shown that the physical mechanisms responsible for the existence of these wavelengths are completely different. Knowledge of the precise spectral location of both the wavelengths are necessary to avoid false positives/negatives in the measurand quantity and to maximize the spectral sensitivity.

\section{Acknowledgement}
The work was financially supported by Science and Engineering Research Board, Government of India through projects PDF/2017/002679 and EMR/2016/007936.

\end{document}